\documentclass[twocolumn,showpacs,amsmath,amssymb,aps,pra,floatfix]{revtex4}
\usepackage{bm}
\usepackage{amsfonts}
\usepackage{graphicx}
\begin{document}

\title{Relative energy dependence of the Bethe-Salpeter amplitude in two-dimensional massless QED}
\author{Tomasz Rado\.zycki}
\email{t.radozycki@uksw.edu.pl} \affiliation{Faculty of Mathematics and Natural Sciences, College of Sciences,
Cardinal Stefan Wyszy\'nski University, W\'oycickiego 1/3, 01-938 Warsaw, Poland}

\begin{abstract}
The exact $q\bar{q}$ Bethe-Salpeter bound state amplitude is investigated in the space of relative energy $E$ for fixed value of relative position. By means of approximate analysis it is shown to possess singularities in $E$ whenever one of the quarks reaches the energy threshold for the creation of certain number of Schwinger bosons. These results are afterwards confirmed by numerical plots.

\end{abstract}
\pacs{11.10.Kk, 11.10.St} 
\maketitle

\section{Introduction}

One of the essential and unsolved problems in quantum field theory is the correct description of bound states in a truly relativistic way, with retardation effects taken into account. On one hand it belongs to the less understood aspects of QFT and on the other it is one of the most important, necessary for the explanation of the whole spectrum of hadrons.  Over sixty years ago the equation for the bound state amplitude -- the so called Bethe-Salpeter (B-S) equation -- was proposed~\cite{bs,gml} but its solutions are still lacking even in model studies. 

Solving the B-S equation is not an easy task: it is an integral equation in several dimensions but, what is worse, it requires from the very beginning the knowledge of the explicit form of the nonperturbative propagators for bound particles, as well as their interaction kernel. Because none of these quantities is known in realistic field theory or even expected to be known, one is inevitably led to strong simplifications as for instance that consisting on replacing full propagators with free ones and using the so called ladder approximation. But even then the problem is still very complicated and has found its solution only in few model calculations~\cite{wick,cut,nn1}. In hadronic physics a step forward has been done in some numerical studies exploiting the truncated (i.e. again approximated) QCD~\cite{bash}.

A novel and troublesome feature of B-S equation is the appearance of the additional variable, absent in nonrelativistic studies: the so called relative time or equivalently the relative energy. There have been undertaken many attempts to simplify the equation by neglecting the relative time dependence of the amplitude~\cite{fg,bb,lusg,lus,bij,lucha,nn0,hs}. Such approaches, however, lead to oversimplification: by imposing the instantaneity condition they throw aside the most important feature, which differentiates the relativistic description from the nonrelativistic one. In such an approach the most interesting part of information is then lost because of the approximation itself. 

The analysis of that nontrivial aspect of a bound system has been undertaken in a couple of works~\cite{keister,ct, bb1,kopa0,kopa,suche,pd} both in euclidean~\cite{zuil,guth,efi,dork2,dork1,carb1,kara} and in Minkowski space~\cite{kara,kara1,kusaka,kusaka1,carb2,carb3,carb4,hall} (where it is more difficult), but a complete description of any B-S amplitude in its full complexity is missing, with one exception to be mentioned below. That has been mainly due to the lack of the appropriate model, which on one hand would constitute a nontrivial quantum field theory and on the other would allow for analytic and exact solutions.  However, a couple of years ago the B-S amplitude in the full analytic form was found in position space in the massless quantum electrodynamics in two dimensions (in the so called Schwinger Model (SM)~\cite{js})~\cite{trsing,trfactor} allowing {\em ipso facto} for more detailed investigation of the role played by relative variables in a nontrivial field theory. 

The model in question is defined through the Lagrangian density in $2D$:
\begin{eqnarray} 
{\cal L}(x)=&&\!\!\!\!\!\!\overline{\Psi}(x)\bigg(i\gamma^{\mu}\partial_{\mu}-
gA^{\mu}(x)\gamma_{\mu}\bigg)\Psi (x)\nonumber\\
&&\!\!\!\!\!\!- \frac{1}{4}F^{\mu\nu}(x)F_{\mu\nu}(x)- \frac{\xi}{2}\left(\partial_{\mu}A^{\mu}(x)\right)^2, 
\label{eq:lagr} 
\end{eqnarray} 
where $g$ is the coupling constant and $\xi$ -- the gauge fixing parameter (later set to be equal to infinity, which corresponds to the Landau gauge). Dirac gamma matrices may be chosen as two-dimensional objects, for instance in the following form:
\begin{eqnarray}
&&\gamma^0=\left(\begin{array}{lr}0 & \hspace*{2ex}1 \\ 1 & 0 
\end{array}\right), \;\;\;\;\; 
\gamma^1=\left(\begin{array}{lr} 0 & -1 \\ 1 & 0 
\end{array}\right),\nonumber\\ 
&&\gamma^5=\gamma^0\gamma^1=\left( 
\begin{array}{lr} 1 & 0 \\ 0 & -1 \end{array}\right),
\label{eq:gammas}
\end{eqnarray}
and for the metric tensor we have $\mathfrak{g}^{00}=-\mathfrak{g}^{11}=1$.
 
For the last fifty years this model has become a perfect testing laboratory for various nonperturbative properties of QFT. The particular attention deserve in this context confinement, topological sectors, instantons and condensates. Among other interesting features one should mention the existence of anomaly and nonzero mass generation of the gauge boson without the necessity of introducing any auxiliary Higgs field. 

In this work we concentrate on another remarkable property of this model, spoken of above. Thanks to the analytic determination of the nonperturbative four-point Green's function~\cite{brown,trjmn,trinst} it allows for the investigation 
of bound states formation. It has been found that the SM model possesses a state, which might be called a `meson', constituting a bound system of a quark and an antiquark (the fundamental fermion of the theory is often called a quark, due to the similarity  between SM and QCD in some respects as for instance confinement) and which is also known as the Schwinger boson of mass $\mu=g/\sqrt{\pi}$. 

In our former papers we concentrated on this issue. 
Firstly, we analyzed the structure of the full four-point (two-fermion) Green's function and found the exact and explicit form of the B-S bound state amplitude in all contributing (i.e. for $k=0,\pm 1$) instanton sectors~\cite{trsing,trfactor}. This result seems to be of certain significance, since, up to our knowledge, it is a unique exact B-S function, which is known exactly in any nontrivial field theory. This result was obtained not by solving the B-S equation, which we are even unable to write down, but by analyzing the residue in the pole corresponding to the Schwinger boson~(cf. \cite{eden,mand}). 

Secondly, the subsequent paper~\cite{trrel} was devoted to the investigation of the properties of the previously found B-S function in the space of relative variables. The present work continues this research concentrating on the relative energy variable. 

The B-S function, which was found from the $t$-channel of two-fermion Green's in the residue of the pole corresponding to $P^2=\mu^2$ (where $P$ is the center-of-mass two-momentum), i.e to the Schwinger boson, has the form:
\begin{equation}
\Phi_P(x)=\Phi_P^{(0)}(x)+\Phi_P^{(1)}(x),
\label{eq:phip}
\end{equation}
where $x=[t,r]$ denotes the relative quark-antiquark coordinate and indices $0$ and $1$ refer to the $k=0$ and $k=\pm 1$ instanton sectors. Higher sectors do not contribute. The Green's function in question is a vacuum (strictly speaking $\theta$-vacuum) expectation value of four fermion fields and the residue factorizes into two bilinear functions. It is well known however, that to change the topological index of the vacuum by $1$ the operator which is quadratic in fermion fields is needed~\cite{rajar,huang}. This is connected with the masslessness of fermions in the theory and the absence of the chirally noninvariant term $m\overline{\Psi}\Psi$ in the Hamiltonian, which results in the suppression of tunneling between topological vacua. Consequently only chirality changing operators can have nonvanishing matrix elements between various topological vacua, with the $2:1$ correspondence between chirality of the operator and instantonic number.

The following expressions for $\Phi_P^{(0)}$ and $\Phi_P^{(1)}$ have been found in the above mentioned works:

\begin{eqnarray}
\Phi_P^{(0)}(x)  =  -2\sqrt{\pi}{\cal S}_0(x)e^{-ig^2\beta(x)}\gamma^5\sin(Px/2)\nonumber\\
  =  C_P^\mu(t,r)\gamma_\mu\gamma^5,\label{eq:phip0}\\
\Phi_P^{(1)}(x)  =  \frac{\mu}{2\sqrt{\pi}}e^{\gamma_E}e^{ig^2\beta(x)}e^{-i\theta\gamma^5}\gamma^5\cos(Px/2)\nonumber\\
  = D_P(t,r)e^{-i\theta\gamma^5}\gamma^5\; ,\label{eq:phip1}
\end{eqnarray}
where $\theta$ is the vacuum parameter. ${\cal S}_0(x)$ denotes here the free fermion propagator:
\begin{equation}
{\cal S}_0(x)=-\frac{1}{2\pi}\ \frac{\not\! x}{x^2-i\varepsilon}
\label{eq:s0}
\end{equation}
and the function $\beta$ appearing in many quantities of the Schwinger Model is defined as  
\begin{eqnarray}  
&&\beta(x)=\label{eq:beta}\\
&&\left\{\begin{array}{l}\frac{i}{2g^2}\left[- 
\frac{i\pi}{2}+\gamma_E +\ln\sqrt{ \mu^2x^2/4}+   
\frac{i\pi}{2}H_0^{(1)}(\sqrt{\mu^2x^2})\right], \\
\hspace*{10ex}
x\;\;\;\; {\rm timelike},\\   
\frac{i}{2g^2}\left[\gamma_E+\ln\sqrt{- 
\mu^2x^2/4}+K_0(\sqrt{-\mu^2x^2})  
\right], \\ \hspace*{10ex} x\;\;\;\; {\rm  
spacelike}.\end{array}\right.  
\nonumber  
\end{eqnarray}  
Symbol $\gamma_E$ is here the Euler constant and $H_0^{(1)}$ and $K_0$ are Hankel function of the first kind and Basset function respectively.

The present paper is organized as follows. In the next section we consider $\Phi_P^{(0)}$, the exact B-S amplitude in the instanton sector $k=0$. We investigate its relative energy dependence for fixed values of relative position in the center-of-mass frame. Using some approximations we identify the location and character of singularities in relative energy. This analysis has qualitative character rather than quantitative. The obtained results are confirmed by the appropriate plots. In section~\ref{sec:1inst} we concentrate on the B-S amplitude in the sector $k=1$. We carry out the same program as in~\ref{sec:0inst} and find similar singularities for the function $\Phi_P^{(1)}$. The last section contains the summary of the obtained results and some conclusions.

\section{The B-S function in the noninstantonic sector}
\label{sec:0inst}

The relative-energy B-S function in the instanton sector $k=0$ is defined through the Fourier transform over relative time:
\begin{equation}
\Phi_P^{(0)}(E,r)=\int\limits_{-\infty}^{\infty} e^{i\frac{\scriptstyle E}{\scriptstyle 2}\,t}\Phi_P^{(0)}(t,r)dt= C_P^\mu(E,r)\gamma_\mu\gamma^5.
\label{eq:f0e}
\end{equation}

Fourier-transformed quantities will be, throughout this paper, identified only by their arguments, without the change of the appropriate symbols. We hope it will not be confusing. The one half in the exponent comes from the definition of the center-of-mass variables and relative ones, we use for the two-body system:
\begin{equation}
X=\frac{1}{2}(x_1+x_2),\;\;\;\; x=x_1-x_2,
\label{eq:relva}
\end{equation}
and hence
\begin{equation}
x_1=X+\frac{1}{2}\,x,\;\;\;\; x_2=X-\frac{1}{2}\,x.
\label{eq:pava}
\end{equation}
With this convention the Fourier exponent in question becomes:
\begin{equation}
e^{i(p_1x_1+p_2x_2)}=e^{iPX+i\frac{\scriptstyle Q}{\scriptstyle 2}\,x}
\label{eq:expf}
\end{equation}
where $P=[P^0,P^1]=p_1+p_2$ is the total (center of mass) two-momentum satisfying $P^2=\mu^2$, and $Q=[E, q^1]=p_1-p_2$ is the relative one. The inverse relations are:
\begin{equation}
p_1=\frac{1}{2}(P+Q),\;\;\;\; p_2=\frac{1}{2}(P-Q)
\label{eq:paqa}
\end{equation}

The quantities $C_P^0(E,r)$ and $C_P^1(E,r)$ may be given the following form:
\begin{eqnarray}
C_P^0(E,r) = A(\omega_+, r)e^{-\frac{i}{2}P^1 r}-A(\omega_-, r)e^{\frac{i}{2}P^1 r},\label{eq:c0e}\\
C_P^1(E,r) = B(\omega_+, r)e^{-\frac{i}{2}P^1 r}-B(\omega_-, r)e^{\frac{i}{2}P^1 r},\label{eq:c1e}
\end{eqnarray}
where we introduced the denotation
\begin{equation}
\omega_\pm=\frac{1}{2}(E\pm P^0).
\label{eq:omegi}
\end{equation}

The coefficient functions $A$ and $B$ after some rearrangements are expressed through the Cauchy principal value integrals as follows:
\begin{eqnarray}
&&\hspace*{-7ex}A(\omega,r)=\nonumber\\ &&\hspace*{-6ex} \frac{1}{\sqrt{\pi}}\,{\cal P}\int\limits_0^\infty dt\, \frac{t \sin \omega t}{t^2-r^2}\,e^{-ig^2\beta(t,r)}+\frac{i\sqrt{\pi}}{2}\,\sin\omega |r|,
\label{eq:aval}\\
&&\hspace*{-7ex}B(\omega,r)=\nonumber\\ &&\hspace*{-6ex} -\frac{ir}{\sqrt{\pi}}\,{\cal P}\int\limits_0^\infty dt\, \frac{\cos \omega t}{t^2-r^2}\,e^{-ig^2\beta(t,r)}+\frac{i\sqrt{\pi}|r|}{2r}\,\cos\omega |r|.
\label{eq:bval}
\end{eqnarray}

From the asymptotics of the Hankel function of the first kind 
it may be easily seen, that for large $t$ the factor $e^{-ig^2\beta(t,r)}$ behaves as:
\begin{equation}
e^{-ig^2\beta(t,r)}\sim \frac{1-i}{2}\,e^{\gamma_E/2} \sqrt{\mu t},
\label{eq:bebe}
\end{equation} 
and, by virtue of the Dirichlet rule, the integral for $A$ owes its convergence in infinity to the interplay between the oscillatory functions $\sin \omega t$ and $H_0^{(1)}$ in~(\ref{eq:beta}). This is an important observation for our conclusions. The convergence of $B$, which has one power of $t$ more in denominator, is absolute, but it may be improved by an oscillatory factor too. However, due to the complicated form of the function $\beta(t,r)$ these integrals cannot be found explicitly and one is doomed to approximate or numerical analysis. It is a common feature of the Schwinger Model, that analytic solutions can analytically be calculated only in coordinate space. It refers to various physical quantities, an exception being purely bosonic Green's functions.

It is clear, that the function $A(\omega,r)$ is odd in the argument $\omega$ and even in relative position $r$, and $B(\omega,r)$ conversely. Consequently $C_P^{0,1}(E, r)$ have the following symmetry properties:
\begin{eqnarray}
C_P^0(-E, r) =  C_P^0(E, -r),\nonumber\\
C_P^1(-E, r) =  -C_P^1(E, -r).
\label{eq:relc12}
\end{eqnarray}
In the center-of-mass frame, where $P^1=0$ and $P^0=\mu$, these quantities satisfy:
\begin{eqnarray}
C_P^0(-E,r)=C_P^0(E,r),&&  C_P^0(E,-r)=C_P^0(E,r),\nonumber\\
\label{eq:cppar1}\\
C_P^1(-E,r)=-C^1(E,r),&&  C_P^1(E,-r)=-C_P^1(E,r).\nonumber\\
\label{eq:cppar2}
\end{eqnarray}

Let us now try to identify the possible  singularities of $A(\omega,r)$ in $\omega$ for fixed $r$. According to what was said above, they arise when the Dirichlet rule fails (or rather may not be applied due to the cancellation among oscillating factors) and therefore it is sufficient to consider the integral from certain point $\lambda |r|$ to infinity (with $\lambda >1$), which will be called $\tilde{A}(\omega,r)$. The last term in~(\ref{eq:aval}) is inessential for this  analysis since it does not contribute to any singularities. Using~(\ref{eq:beta}), we have:
\begin{eqnarray}
\tilde{A}(\omega,r) &= &\frac{1-i}{2}\,\sqrt{\frac{\mu}{\pi}}\,e^{\gamma_E/2}\int\limits_{\lambda|r|}^\infty dt\, \frac{t \sin \omega t}{(t^2-r^2)^{3/4}}\nonumber\\
&& \times \exp\left[i\frac{\pi}{4}\,H_0^{(1)}(\mu\sqrt{t^2-r^2})\right].
\label{eq:tia}
\end{eqnarray}

Since for large values of $t$ the Hankel function behaves as~\cite{gr}
\begin{eqnarray}
&&\hspace*{-4ex}H_0^{(1)}(\mu\sqrt{t^2-r^2}) \sim e^{-i\pi/4}\exp\left[ i\mu\sqrt{t^2-r^2}\right]\nonumber\\
&&\hspace*{-1ex}\times \sqrt{\frac{2}{\pi\sqrt{t^2-r^2}}}\left(1-\frac{i}{8\mu\sqrt{t^2-r^2}}+\ldots\right),
\label{eq:haex}
\end{eqnarray}
each following term of the above expansion as well as that of the exponent function in the integrand of~(\ref{eq:tia}) has better convergence properties, contributing thereby less to the singularities in $\omega$. Therefore the strongest ones, which can eventually be clearly identified on the plots, come only from first few terms:
$$
\tilde{A}(\omega,r)=\tilde{A}_a(\omega,r)+\tilde{A}_b(\omega,r)+\tilde{A}_c(\omega,r)+\ldots
$$
Below we analyze them starting from the most singular one:
\begin{equation}
\tilde{A}_a(\omega,r)=\frac{1-i}{2}\,\sqrt{\frac{\mu}{\pi}}\,e^{\gamma_E/2}\int\limits_{\lambda|r|}^\infty dt\, \frac{t \sin \omega t}{(t^2-r^2)^{3/4}}
\label{eq:tila}
\end{equation}
For large values of $\lambda$ one can neglect $r^2$ in denominator, and the value of the integral can be found to be:
\begin{eqnarray}
\tilde{A}_a(\omega,r)& \approx & \frac{(1-i)\sqrt{\mu}}{2\sqrt{2}}\, e^{\gamma_E/2} \frac{\mathrm{sgn}\, (\omega)}{\sqrt{|\omega|}}\nonumber\\
&&\times\left[1-2S(\sqrt{2\lambda |\omega r|/\pi})\right],
\label{eq:a2a1}
\end{eqnarray}
where $S(x)$ is the Fresnel integral:
\begin{equation}
S(\sigma)=\int\limits_0^\sigma d\tau \sin (\frac{\pi}{2}\, \tau^2).
\label{eq:frei}
\end{equation}

We are interested in the behavior at $\omega\approx 0$ (it is the only possible singular point) with large but fixed $\lambda$, which corresponds to small value of $\sigma$. In such a case we can use the expansion:
\begin{equation}
S(\sigma)\approx \frac{\pi}{6}\, \sigma^3+O(\sigma^4).
\label{eq:sapp}
\end{equation} 
and the first two terms of~(\ref{eq:a2a1}) become:
\begin{eqnarray}
\tilde{A}_a(\omega,r)  &\approx & \frac{(1-i)\sqrt{\mu}}{2\sqrt{2}}\, e^{\gamma_E/2} \frac{\mathrm{sgn}\, (\omega)}{\sqrt{|\omega|}}\nonumber\\&&-\frac{(1-i)\sqrt{\mu}}{3\sqrt{\pi}}\, e^{\gamma_E/2}(\lambda |r|)^{3/2}\omega,
\label{eq:aasi}
\end{eqnarray}

The quantity $\tilde{A}_a$ is then seen to possess the strong singularity at $\omega=0$, which should be represented with an infinite peak on the plot of $\Phi_P^{(0)}(E,r)$.

The subsequent term of~(\ref{eq:tia}) after some rearrangements has the form:
\begin{equation}
\tilde{A}_b(\omega,r)=\frac{1}{4}\, e^{\gamma_E/2}\int\limits_{\lambda|r|}^\infty dt\, \frac{t \sin \omega t}{t^2-r^2}\, \exp[i\mu\sqrt{t^2-r^2}].
\label{eq:tib}
\end{equation}
Again $r^2$ may be omitted for large $\lambda$ and one can obtain for $\tilde{A}_b$ the approximated result:
\begin{equation}
\tilde{A}_b(\omega,r)\approx\frac{i}{8}\, e^{\gamma_E/2}\left[\mathrm{Ei}\, (i\lambda |r|(\mu+\omega)-\mathrm{Ei}\, (i\lambda |r|(\mu-\omega)\right],
\label{eq:tibapp}
\end{equation}
where $\mathrm{Ei}$ is the exponential integral function. It has the following expansion for an imaginary argument:
\begin{equation}
\mathrm{Ei}\, (i\sigma)\approx \ln \sigma+\frac{i\pi}{2}+\gamma_E+i\sigma+O(\sigma^2),
\label{eq:eiex}
\end{equation}
which shows, that $\tilde{A}_b(\omega,r)$ contains the logarithmic singularities at $\omega=\pm\mu$. They should again manifest themselves as infinite peaks, although less pronounced than those of~(\ref{eq:aasi}).

The next term may be given the form:
\begin{eqnarray}
&&\hspace*{-5ex}\tilde{A}_c(\omega,r)=\frac{(1+i)\sqrt{\pi}}{16\sqrt{\mu}}\, e^{\gamma_E/2}\nonumber\\
&&\times\int\limits_{\lambda|r|}^\infty dt\, \frac{t \sin \omega t}{(t^2-r^2)^{5/4}}\, \exp[2i\mu\sqrt{t^2-r^2}].
\label{eq:tic}
\end{eqnarray}
One can show in the same way as above, that it has a branch points of the kind $\sqrt{\omega\pm 2\mu}$, but the integral is convergent even for $\omega=\pm 2\mu$

The quantity $B$ is smoother, but it also contributes to singularities. We define $\tilde{B}$ in an analogous way to $\tilde{A}$ and find that the most singular term is:
\begin{equation}
\tilde{B}_a(\omega,r)\approx-r\frac{(1+i)\sqrt{\pi}}{2\sqrt{\mu}}\, e^{\gamma_E/2}\int\limits_{\lambda|r|}^\infty dt\, \frac{ \cos \omega t}{(t^2-r^2)^{3/4}}.
\label{eq:biba}
\end{equation}
This integral is convergent in an absolute way and it may be shown to contain a singularity of the kind $\sqrt{|\omega|}$, which might eventually appear in the form of slight cusps, but they will be hidden by the peaks of $\tilde{A}_a$ in the same place. The following terms in $\tilde{B}(\omega,r))$ need not be considered.

Because $\Phi_P^{(0)}$ is a combination of $A(\omega_+,r)$ and $A(\omega_-,r)$ via the formulae~(\ref{eq:f0e}), (\ref{eq:c0e}) and~(\ref{eq:c1e}), we expect the manifestations of the following well-defined singularities in the form of peaks in the B-S amplitude:
\begin{enumerate}
\item $\omega_\pm=0\;\;\; \Longrightarrow \;\;\; E=\pm\mu$.
\item $\omega_\pm=\mu\;\;\; \Longrightarrow \;\;\; E=\mu,\, E=3\mu$.
\item $\omega_\pm=-\mu\;\;\; \Longrightarrow \;\;\; E=-\mu,\, E=-3\mu$.
\end{enumerate}
The singularities coming from~(\ref{eq:tic}) occurring at $\omega_\pm=2\mu$ and $\omega_\pm=-2\mu$, which correspond to $E=-5\mu, -3\mu, 3\mu, 5\mu$ are expected to be less visible. What is important, the positions of all the singularities are not affected by the value of the relative position $r$. These analytical although approximate results are confirmed below by the appropriate numerical graphs.

Since $\Phi_P^{(0)}$ is a \mbox{$2\times 2$} complex matrix, it cannot be directly represented on the plot. To show the relative energy dependence of the B-S amplitude we, therefore, define a scalar, similarly as it was done in~\cite{trrel}:
\begin{equation}
|\Phi_P^{(0)}|= \left(\frac{1}{2}\ \mathrm{tr}[\Phi_P^{(0)+}\Phi_P^{(0)}]\right)^{1/2}.
\label{eq:modp1}
\end{equation}
It is a gauge invariant, dimensionless quantity and can serve as a measure for the strength of the B-S amplitude. It may be simply expressed in terms of $A$ and $B$, as
\begin{eqnarray}
|\Phi_P^{(0)}(E,r)|&&\!\!\!= \left\{|C_P^0(E,r)|^2+|C_P^1(E,r)|^2\right\}^{1/2}=\nonumber\\
&&\!\!\!\!\!\!\big\{[\bar{A}(\omega_+, r)e^{\frac{i}{2}P^1 r}-\bar{A}(\omega_-, r)e^{-\frac{i}{2}P^1 r}]\nonumber\\ &&\!\!\!\!\!\!\times [A(\omega_+, r)e^{-\frac{i}{2}P^1 r}-A(\omega_-, r)e^{\frac{i}{2}P^1 r}]\nonumber\\
&&\!\!\!\!\!\! + [\bar{B}(\omega_+, r)e^{\frac{i}{2}P^1 r}-\bar{B}(\omega_-, r)e^{-\frac{i}{2}P^1 r}]\nonumber\\ &&\!\!\!\!\!\!\times  [B(\omega_+, r)e^{-\frac{i}{2}P^1 r}-B(\omega_-, r)e^{\frac{i}{2}P^1 r}]\big\}^{1/2},\nonumber\\
\label{eq:piab}
\end{eqnarray}
where bars over symbols refer to the complex conjugation.

To investigate the relative energy dependence of the B-S function, the most natural is to perform the plot of $|\Phi_P^{(0)}(E,r)|$ in the CM frame, putting thereby $P^1=0$. In this frame it takes the form:
\begin{eqnarray}
|\Phi_P^{(0)}(E,r)|&&\!\!\!= \nonumber\\
&&\hspace*{-9ex}\big\{[\bar{A}(\omega_+, r)-\bar{A}(\omega_-, r)]\,[A(\omega_+, r)-A(\omega_-, r)]\nonumber\\
&&\hspace*{-9ex} + [\bar{B}(\omega_+, r)-\bar{B}(\omega_-, r)]\, [B(\omega_+, r)-B(\omega_-, r)]\big\}^{1/2}.\nonumber\\
\label{eq:piaba}
\end{eqnarray}
and becomes an even function in both relative energy and relative position. In a boosted frame the plotted function is no longer even, but the positions of singularities in relative energy remain invariant. 

\begin{figure*}
\centering
{\includegraphics[width=1\textwidth]{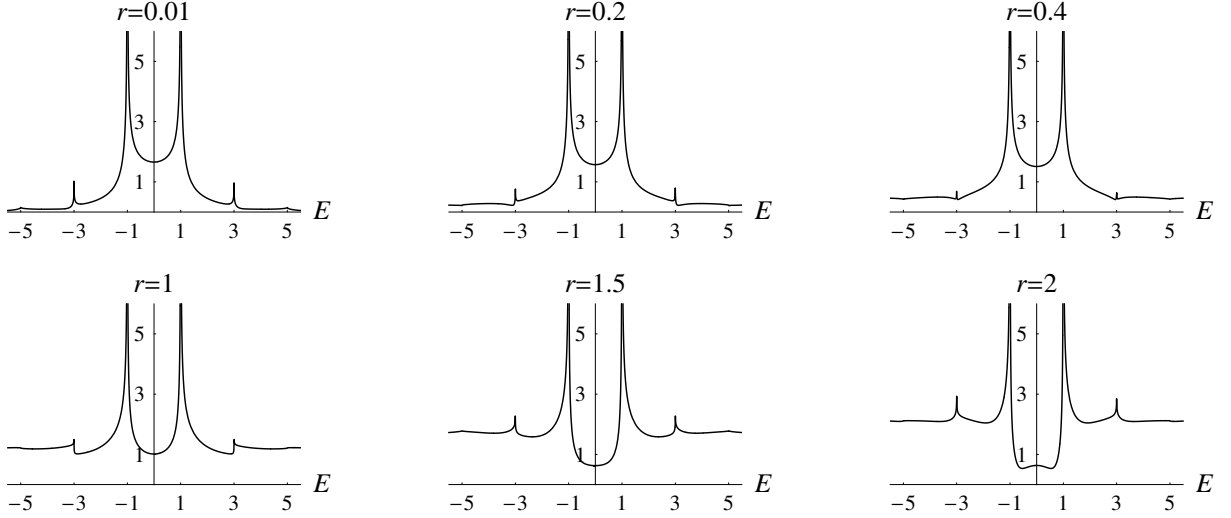}
\caption{The dependence of the amplitude $|\Phi_P^{(0)}|$ on  relative energy $E$ with fixed relative position $r$. The variable $r$ for the successive plots is chosen to be: $0.01$, $0.2$, $0.4$, $1$, $1.5$, $2$. $E$ is measured in units of $\mu$ and $r$ in units of $\mu^{-1}$.}
\label{fig:plr-E}}
\end{figure*}

Figure~\ref{fig:plr-E} shows the dependence of~(\ref{eq:piaba}) on relative energy for certain fixed relative positions. The main noticeable feature is the presence of strong peaks for the relative energy constituting an odd multiple of the Schwinger mass $\mu$. This is what we expected from our approximations. 

Some explanation require the behavior at $E=\pm 3\mu$, where, according to our former analysis, the function should display the logarithmic nature. This ought to lead to infinite peaks, and this is really the case, when investigated carefully. It may be, however, estimated that they are extremely narrow. For instance at the value of $1$ the width becomes almost $10$ times narrower than the thickness of the plotting line, and therefore it cannot be properly represented on the drawing. The singularities at $E=\pm5\mu$ can also be identified, but it requires very high resolution plots.

The appearance of the singularities at relative energies corresponding to an odd number of Schwinger boson masses $\mu$ may be understood from the simple picture. Let us consider a quark-antiquark pair of energies $E_1$ and $E_2$ which form a bound state, i.e. a meson of the CM energy equal to $\mu$. If, due to some fluctuation, one quark acquires the energy sufficient to create a meson, i.e. $E_1=\mu$, becoming of course a deeply off-shell particle (if one at all can talk about on-shell particles in the bound state) then we must have $E_2=0$ and the relative energy $E=E_1-E_2=\mu$. This corresponds to the first two singularities.

Now assume that a quark gained the energy sufficient to create $n$ bosons. Then $E_1=n\mu$ and $E_2=(1-n)\mu$ in order that $E_1+E_2=\mu$ (in this rough image). In such a case for the relative energy we obtain: $E=E_1-E_2=(2n+1)\mu$. Such fluctuations are however much less probable and hence higher singularities are more soft.

The other observation, which can be done by analyzing the $E$ and $r$ dependence of $|\Phi_P^{(0)}|$ is that for small relative energies the amplitude decreases with increasing $r$. However for $E$ exceeding the first threshold, i.e for $E>\mu$, the described behavior becomes opposite: the amplitude grows with relative distance. This observation has yet some limitations: we were not able to verify it for very big values of $r$ since the oscillatory integrals~(\ref{eq:aval}) and~(\ref{eq:bval}) become then extremely slowly convergent. As we will see in the following section, this behavior will not be the same for the $k=\pm 1$ instanton sector. 

\section{Instantonic contribution to the B-S function}
\label{sec:1inst}

The relative-energy B-S function in the instanton sector $k=\pm 1$ has the form:
\begin{equation}
\Phi_P^{(1)}(E,r)=\int\limits_{-\infty}^{\infty} e^{i\frac{\scriptstyle E}{\scriptstyle 2}\,t}\Phi_P^{(1)}(t,r)dt= D_P(E,r)e^{-i\theta\gamma^5}\gamma^5.
\label{eq:f1e}
\end{equation}

The tensor structure of $\Phi_P^{(1)}$ is simpler than that of $\Phi_P^{(0)}$. It contains only one (scalar) coefficient $D_P(E,r)$, which may be written in terms of a new function $F(\omega,r)$ defined as:
\begin{equation}
F(\omega,r)=\frac{\mu}{2\sqrt{\pi}}\,e^{\gamma_E}\int\limits_0^\infty dt\cos \omega t\, e^{ig^2\beta(t,r)},
\label{eq:deff}
\end{equation}
in the form
\begin{equation}
D_P(E,r)=F(\omega_+, r)e^{-\frac{i}{2}P^1 r}-F(\omega_-, r)e^{\frac{i}{2}P^1 r}.\label{eq:d1e}
\end{equation}

For large $t$  the exponent under the integral has the following behavior:
\begin{equation}
e^{ig^2\beta(t,r)}\sim (1+i)e^{-\gamma_E/2}\,\frac{1}{\sqrt{\mu t}},
\label{eq:bebea}
\end{equation}
which again guarantees the convergence (via Dirichlet rule) of the improper integral~(\ref{eq:deff}) except for the certain particular values of $\omega$ corresponding to divergent singularities. 

The function $F(\omega, r)$ is, in an obvious way, an even function in both arguments, which gives for $D_P$:
\begin{equation}
D_P(-E,r)=D_P(E,-r).
\label{eq:symd}
\end{equation}
Consequently in the CM frame $D_P(E,r)$ has the following symmetry properties:
\begin{equation}
D_P(-E,r)=-D_P(E,r),\;\;\;\;  D_P(E,-r)=D_P(E,r),
\label{eq:symdcm}
\end{equation}

Let us now concentrate on the possible singularities of  $F(\omega,r)$ as the function of $\omega$ with fixed value of relative position $r$. Similarly as it was for the quantity $A(\omega,r)$ we focus on the behavior of the integrand function at infinity. We therefore define:
\begin{eqnarray}
\tilde{F}(\omega,r) &=& \frac{1+i}{2}\,\sqrt{\frac{\mu}{\pi}}\,e^{\gamma_E/2}\int\limits_{\lambda|r|}^\infty dt\, \frac{\cos \omega t}{(t^2-r^2)^{1/4}}\nonumber\\
&&\!\!\!\times \exp\left[-i\frac{\pi}{4}\,H_0^{(1)}(\mu\sqrt{t^2-r^2})\right],
\label{eq:tid}
\end{eqnarray}
assuming that the parameter $\lambda$ is large. Using again the expansion~(\ref{eq:haex}) we find the less convergent term in the form:
\begin{equation}
\tilde{F}_a(\omega,r)=\frac{1+i}{2}\,\sqrt{\frac{\mu}{\pi}}\,e^{\gamma_E/2}\int\limits_{\lambda|r|}^\infty dt\, \frac{\cos \omega t}{(t^2-r^2)^{1/4}}
\label{eq:tild}
\end{equation}

For $\lambda \gg 1$ one can omit $r^2$ as before, and the approximate value of the above integral is:
\begin{eqnarray}
\tilde{F}_a(\omega,r)&  \approx &  \frac{(1+i)\sqrt{\mu}}{2\sqrt{2}}\, e^{\gamma_E/2}\, \frac{\mathrm{sgn}\, (\omega)}{\sqrt{|\omega|}}\nonumber\\
&&\times\left[1-2C(\sqrt{2\lambda |\omega r|/\pi})\right],
\label{eq:d2a1}
\end{eqnarray}
where $C(x)$ is the second Fresnel integral, defined as:
\begin{equation}
C(\sigma)=\int\limits_0^\sigma d\tau \cos (\frac{\pi}{2}\, \tau^2).
\label{eq:freic}
\end{equation}
For small value of $\sigma$ it may be approximated, according to the known formula:
\begin{equation}
C(\sigma)\approx \sigma -\frac{\pi^2}{40}\, \sigma^5+O(\sigma^6),
\label{eq:capp}
\end{equation} 
and we find in $\tilde{F}_a(\omega,r)$ the similar singularity at $\omega=0$ as it was identified in $\tilde{A}_a(\omega,r)$:
\begin{equation}
\tilde{F}_a(\omega,r)\approx \frac{(1+i)\sqrt{\mu}}{2\sqrt{2}}\, e^{\gamma_E/2}\, \frac{\mathrm{sgn}\, (\omega)}{\sqrt{|\omega|}}.
\label{eq:dasi}
\end{equation}
 
\begin{figure*}
\centering
{\includegraphics[width=1\textwidth]{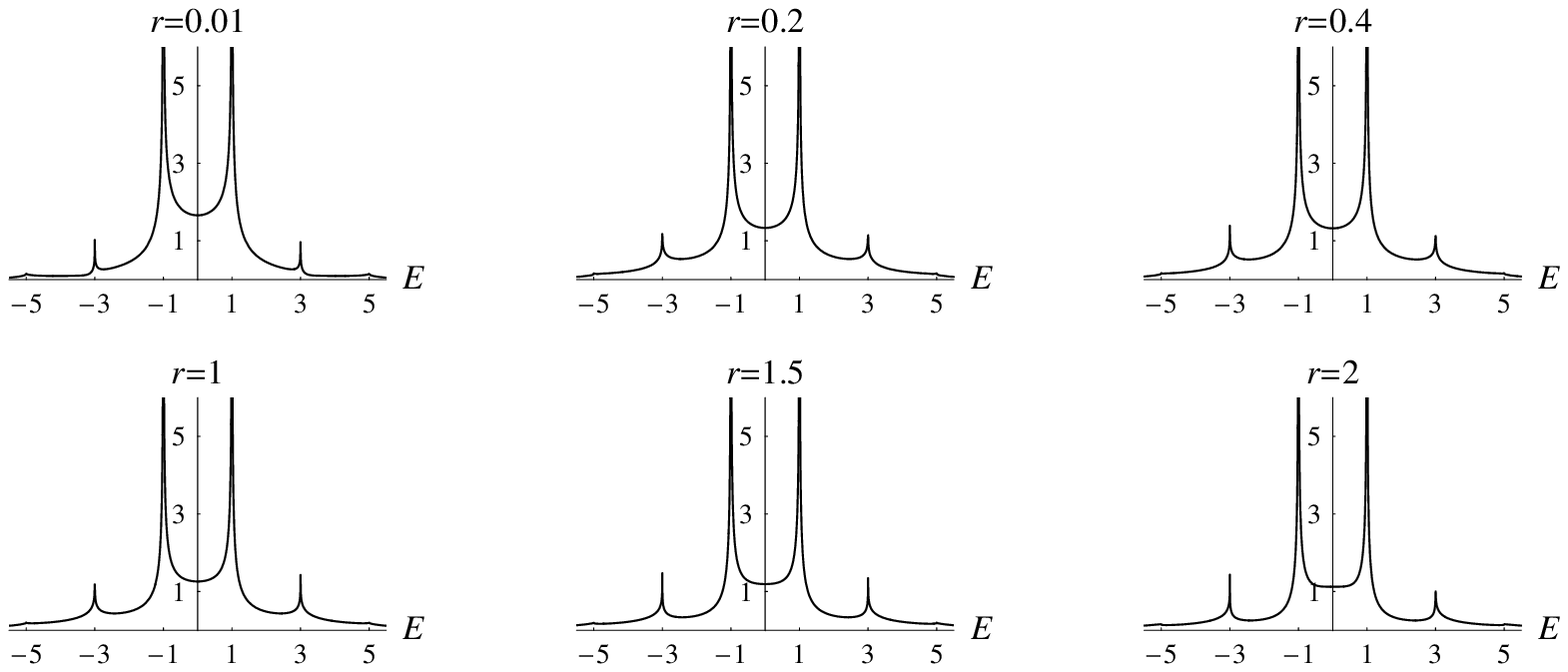}
\caption{Same as Figure~\ref{fig:plr-E}, but for $|\Phi_P^{(1)}|$.} \label{fig:plr1-E}}
\end{figure*}

For the next term of the expansion of~(\ref{eq:tid}) we obtain:
\begin{equation}
\tilde{F}_b(\omega,r)=-\frac{i}{4}\, e^{\gamma_E/2}\int\limits_{\lambda|r|}^\infty dt\, \frac{\cos \omega t}{(t^2-r^2)^{1/2}}\, \exp[i\mu\sqrt{t^2-r^2}],
\label{eq:tfib}
\end{equation}
and after the standard approximation it becomes:
\begin{equation}
\tilde{F}_b(\omega,r)\approx\frac{i}{8}\, e^{\gamma_E/2}\left[\mathrm{Ei}\, (i\lambda |r|(\mu+\omega)+\mathrm{Ei}\, (i\lambda |r|(\mu-\omega)\right].
\label{eq:tfibapp}
\end{equation}
Using~(\ref{eq:eiex}), we see that~(\ref{eq:tfibapp}) is logarithmically divergent as $\omega\rightarrow\pm\mu$. It is the identical result as in the case of $\tilde{A}_b(\omega,r)$.

The last term we are interested in, is
\begin{eqnarray}
&&\hspace*{-5ex}\tilde{F}_c(\omega,r)=-\frac{(1-i)\sqrt{\pi}}{16\sqrt{\mu}}\, e^{\gamma_E/2}\nonumber\\
&&\times\int\limits_{\lambda|r|}^\infty dt\, \frac{\cos \omega t}{(t^2-r^2)^{3/4}}\, \exp[2i\mu\sqrt{t^2-r^2}].
\label{eq:tfic}
\end{eqnarray}
Due to sufficiently high power of $t$ in denominator, this integral is convergent even for $\omega=\pm 2\mu$. It may be shown to contain the slight singularity of the kind $\sqrt{\omega\pm 2\mu}$.

One can then summarize, that the instanton sector $k=\pm 1$ reveals identical singularities at the same values of $\omega$ as for $k=0$. It is also confirmed by the appropriate plots. 

Similarly to~(\ref{eq:modp1}) we define now the quantity, which represents the `value' of the B-S function in this sector:
\begin{equation}
|\Phi_P^{(1)}|= \left(\frac{1}{2}\ \mathrm{tr}[\Phi_P^{(1)+}\Phi_P^{(1)}]\right)^{1/2}.
\label{eq:modp2}
\end{equation}
It is independent both on the choice of gauge and on the vacuum $\theta$ parameter and may be given the form:
\begin{eqnarray}
&&\!\!\!\!\!\!\!\!\!\!|\Phi_P^{(1)}(E,r)|= |D_P(E,r)|=\nonumber\\
&&\big\{[\bar{F}(\omega_+, r)e^{\frac{i}{2}P^1 r}-\bar{F}(\omega_-, r)e^{-\frac{i}{2}P^1 r}]\nonumber\\ &&\times [F(\omega_+, r)e^{-\frac{i}{2}P^1 r}-F(\omega_-, r)e^{\frac{i}{2}P^1 r}]\big\}^{1/2},
\label{eq:pif}
\end{eqnarray}
which in the CM frame reduces to the simple expression:
\begin{eqnarray}
|\Phi_P^{(1)}(E,r)|&&\!\!\!= \nonumber\\
&&\hspace*{-9ex}\big\{[\bar{F}(\omega_+, r)-\bar{F}(\omega_-, r)]\,[F(\omega_+, r)-F(\omega_-, r)]\big\}^{1/2},\nonumber\\
\label{eq:pifa}
\end{eqnarray}

Similarly to $|\Phi_P^{(0)}(E,r)|$ it becomes an even function of both relative position and relative energy.
In figure~\ref{fig:plr1-E} the plot of~(\ref{eq:pifa}) is presented as a function of $E$ for various fixed values of $r$.

The general  relative energy dependence reflects the principal features of the trivial topological sector. The peaks at $E=(2n+1)\mu$ for small values of $n$ are clearly visible. However, contrary to Fig.~\ref{fig:plr-E}, there is no apparent enhancement with increasing $r$ for relative energy exceeding the first threshold. 

On the following figure we plot for completeness the dependence on $E$ of the full B-S function $|\Phi_P|$ defined as:
\begin{equation}
|\Phi_P|= \sqrt{\frac{1}{2}\ \mathrm{tr}[\Phi_P^{+}\Phi_P]}=\sqrt{|\Phi_P^{(0)}|^2+|\Phi_P^{(1)}|^2}.
\label{eq:topi}
\end{equation}

\section{Summary}
\label{sec:sum}

In the summary one should emphasize, that the Schwinger Model turns out to be an exceptional tool to investigate the properties of the bound states in quantum field theory (apart from another nonperturbative aspects).  The fact that the exact form of the B-S amplitude is explicitly known constitutes a unique opportunity to analyze its dependence on relative variables and particularly the role played by the relative energy. The results obtained in the present work show that the `strength' of the B-S function defined by formulae~(\ref{eq:modp1}) and~(\ref{eq:modp2}) exhibits singularities for odd number of Schwinger boson mass $\mu$. The appearance of this threshold structure remains in some correspondence with the results of the approximated scalar-scalar model~\cite{keister}. This effect finds its justification in a simple picture, in which we attribute energies $E_1$ and $E_2$ to both quarks and keep fixed the total energy, which in the center-of-mass frame is equal to $\mu$. The location of singularities is identical in all instanton sectors. 

The other effect, present only in the instanton sector $k=0$ is the enhancement of the `value' of the B-S function for medium values of $r$ (in units of $\mu^{-1}$) for relative energy exceeding the first threshold for production of Schwinger bosons, and opposite behavior for small $r$. The region of large relative positions was not investigated due to obstacles of numerical origin.

Our formulae show, that in a boosted frame, the distribution of singularities remains unaltered, although plots loose in this case their symmetric character with respect to the replacement $E\mapsto -E$.  

It should be noted, that the results of sections~\ref{sec:0inst} and~\ref{sec:1inst} stay in agreement with those of~\cite{trrel}, where it was found, that the singularities in the two-momentum space are located on hyperbolas (see formulae~(17)). In the CM frame and in the notation of the present work, they have the form:
\begin{eqnarray}
\frac{1}{4}(E+\mu)^2-(q^1)^2 =  n^2\mu^2,\nonumber\\
\frac{1}{4}(E-\mu)^2-(q^1)^2 =  n^2\mu^2,
\label{eq:hyp}
\end{eqnarray}
and may be rewritten as
\begin{eqnarray}
q^1 &= & \pm\frac{1}{2}\,\sqrt{(E+\mu)^2-4n^2\mu^2}\nonumber\\
& = & \pm\frac{1}{2}\,\sqrt{(E-(2n-1)\mu)(E+(2n+1)\mu)},\label{eq:hyp1a}\\
q^1& = & \pm\frac{1}{2}\,\sqrt{(E-\mu)^2-4n^2\mu^2}\nonumber\\
& = & \pm\frac{1}{2}\,\sqrt{(E+(2n-1)\mu)(E-(2n+1)\mu)}.
\label{eq:hyp1b}
\end{eqnarray}
   
\begin{figure*}
\centering
{\includegraphics[width=1\textwidth]{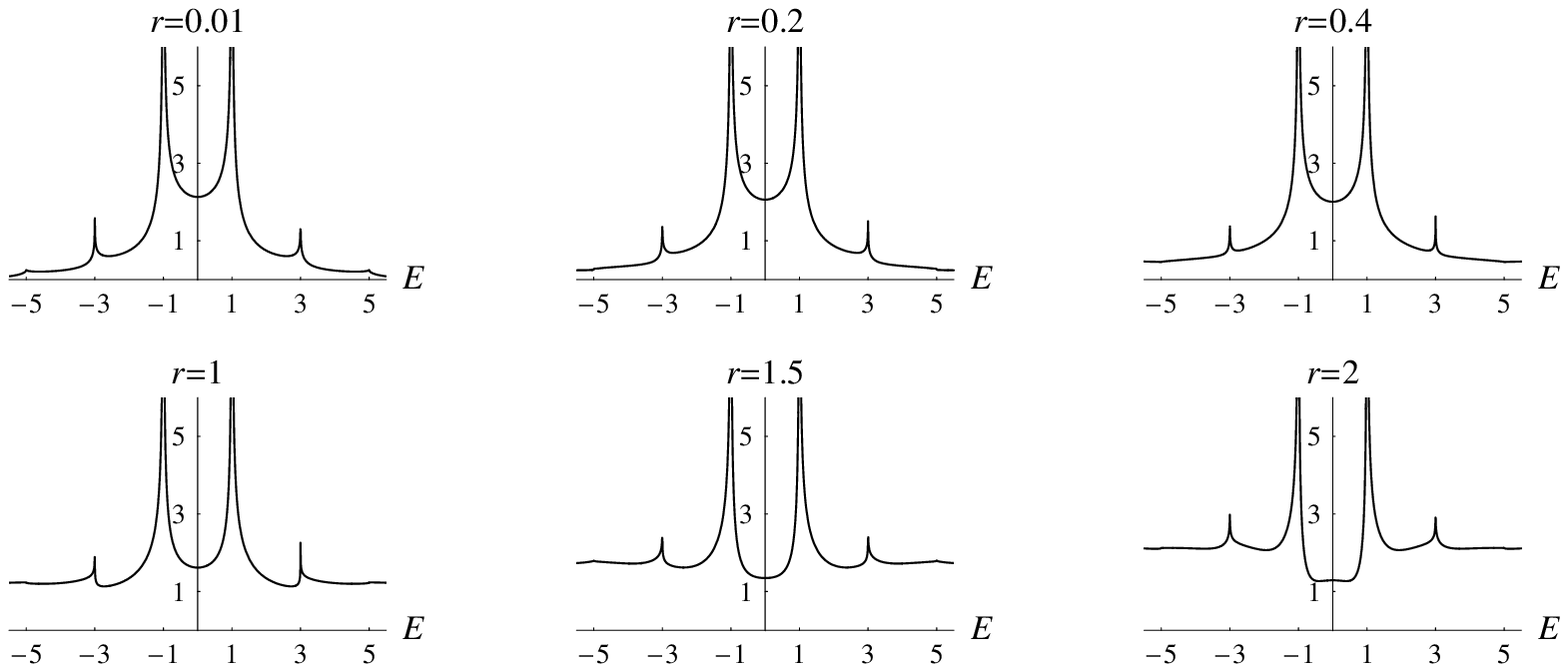}
\caption{Same as Figure~\ref{fig:plr-E}, but for $|\Phi_P|$.} \label{fig:plr2-E}}
\end{figure*}

Now, in the Fourier integral over $q^1$ the singularities in $E$ arise, according to the Landau procedure~\cite{landau,eden,elop}, when two singularities of the integrand functions merge. This happens for $E=(2n-1)\mu$ or $E=-(2n+1)\mu$ in the first case, and for $E=-(2n-1)\mu$ or $E=(2n+1)\mu$ in the second one. They all correspond to an odd number of Schwinger boson masses, as found in the present work. This is only a rough argumentation, since the explicit form on the B-S amplitude in two-momentum space is unknown.

The results obtained here may be of some interest for studies of bound states in more realistic models both in particle and nuclear physics. It is believed that the retardation effects may play an essential role in true QFT bound states~\cite{bha1}. One should also mention, that processes in which bound states appear in internal lines require Feynman integrations over relative coordinates or momenta and are strongly affected by the presence of singularities. The B-S function appears also in the matrix elements and scattering processes involving bound states~\cite{mand, hua} and its role is significant to properly describe the meson decay properties. Therefore any insight into its structure seems to deserve particular attention.

\end{document}